\documentclass[article]{aastex63}

\newcommand{\msun}{M\hbox{$_\odot$}}

\newcommand{\hi}{\ion{H}{1} }

\received{April 13, 2022}
\revised{June 22, 2022}
\accepted{August 3, 2022}

\submitjournal{ApJ}

\shorttitle{Radio emission in Minkowski's Object}
\shortauthors{Nolting et al.}

\graphicspath{{./}{figures/}}

\usepackage{amsmath}
\usepackage{enumitem}
\usepackage{xcolor}
\usepackage{hyperref}

\begin{document}

\title{Observations and Simulations of Radio Emission and Magnetic Fields in Minkowski's Object}

\correspondingauthor{Chris Nolting}
\email{noltingcr@cofc.edu}

\author[0000-0002-0775-6017]{C. Nolting}
\affiliation{Department of Physics and Astronomy, College of Charleston, 66 George Street, Charleston, SC 29424, USA}

\author{M. Lacy}
\affiliation{National Radio Astronomy Observatory, 520 Edgemont Road, Charlottesville, VA 22903, USA}

\author[0000-0003-4823-129X]{S. Croft}
\affiliation{Department of Astronomy, University of California at Berkeley, 501 Campbell Hall, Berkeley, CA 94720, USA}
\affil{Eureka Scientific, 2452 Delmer Street, Suite 100, Oakland, CA 94602-3017, USA}

\author[0000-0002-5786-186X]{P. C. Fragile}
\affiliation{Department of Physics and Astronomy, College of Charleston, 66 George Street, Charleston, SC 29424, USA}
\affil{Kavli Institute for Theoretical Physics, University of California Santa Barbara, Santa Barbara, CA 93106, USA}

\author{S. T. Linden}
\affiliation{Department of Astronomy, University of Virginia, 530 McCormick Road, Charlottesville, VA 22904, USA}

\author[0000-0003-1991-370X]{K. Nyland}
\affiliation{U.S. Naval Research Laboratory, 4555 Overlook Ave SW, Washington, DC 20375, USA}

\author[0000-0002-9471-8499]{P. Patil}
\affiliation{National Radio Astronomy Observatory, 530 McCormick Road
Charlottesville, VA 22904, USA}

\begin{abstract}
We combine new data from the Karl G.\ Jansky Very Large Array with previous radio observations to create a more complete picture of the ongoing interactions between the radio jet from galaxy NGC\,541 and the star-forming system known as Minkowski's Object (MO). We then compare those observations with synthetic radio data generated from a new set of magnetohydrodynamic simulations of a jet-cloud interaction specifically tailored to the parameters of MO. The combination of radio intensity, polarization, and spectral index measurements all convincingly support the interaction scenario and provide additional constraints on the local dynamical state of the intracluster medium and the time since the jet-cloud interaction first began. In particular, we show that only a simulation with a bent radio jet can reproduce the observations.

\end{abstract}

\keywords{galaxies:jets --- galaxies:intracluster medium --- magnetohydrodynamics (MHD) --- galaxies:individual(NGC541) --- galaxies:individual(Minkowski's Object) ---  radio continuum: galaxies}

\section{Introduction} \label{sec:intro}

Feedback from jets produced by active galactic nuclei (AGN) plays a key role in regulating galaxy growth \citep[e.g.][]{croton:06,2012ARA&A..50..455F,2021MNRAS.508.4738M}, heating gas in the interstellar and circumstellar media of galaxies and suppressing star formation on long timescales. There is increasing evidence, however, that interactions between radio jets and denser, clumpy gas, may also {\em increase} star formation rates (particularly on shorter timescales) by an order of magnitude or more \citep{bieri:16}. Such jet-induced star formation (JISF) may play a significant role in producing stars in the early Universe \citep{mdb:08,gaibler:12}. Observations and simulations support a picture in which jets drive radiative shocks in overdense clouds in the multiphase ISM, triggering gravitational collapse \citep{silk:09}. The balance of negative and positive feedback may fine-tune star formation rates over cosmic history, and help explain the correlation between the masses of black holes and their host galaxies.

Although JISF is well-supported by simulations \citep{gaibler:12,mukherjee+18,mukherjee+21} and observations \citep{mccarthy:87,cmvb:87,salome+15,nesvadba+20}, and despite its importance in understanding galaxy formation, only a handful of systems have been clearly identified and studied in this context. One of the best examples is the $z=0.019$ ``Minkowski's Object'' (MO) in Abell~194, where a burst of star formation is being triggered by the radio jet from the nearby active galaxy NGC\,541 \citep[][hereafter C06]{wvb:85,croft:06}. Unlike other local examples of JISF --- for example, Centaurus A \citep{2000ApJ...536..266M} and Henize 2-10 \citep{2022Natur.601..329S} --- the JISF region in MO is well outside the AGN host galaxy, making analysis more straightforward. The UV morphology of MO along the radio jet, as presented by C06, resembles the radio-aligned, rest-frame UV morphologies seen in many high-$z$ radio galaxies, also thought to be caused by JISF \citep[e.g.][]{Dey97}.

Fitting stellar synthesis models to integrated photometry of MO by C06 yielded an age of $7.47 \pm 0.12$\,Myr for the dominant stellar population. More recently, \citet{2020MNRAS.499.4940Z} presented integral field spectroscopy of MO, confirming the JISF scenario and obtaining stellar ages in agreement with C06. \citet{lacy:alma} obtained molecular line and continuum observations of MO using the Atacama Large Millimeter Array (ALMA). Their analysis clearly shows a reduction in jet collimation in the interaction region, and synchrotron emitting material mingling with the ISM in the galaxy. They found that star formation is highly efficient in the upstream regions of MO that were struck first by the jet, and less efficient downstream.

\citet{mo_sim} simulated the collision of a jet with a gas cloud analogous to MO, and found that the star formation could be explained by the driving of a radiative shock into the cloud, causing cloudlets to compress, radiatively cool, and break up into numerous dense, cold fragments. The fragments survive for many dynamical timescales and are presumably precursors to star formation. This motivated follow-up \hi observations (C06) with the Karl G. Jansky Very Large Array (VLA) and GALEX UV observations of MO, and to reanalyze archival VLA continuum and Hubble Space Telescope (HST) data. C06 discovered a double \hi cloud with a mass of $4.9 \times 10^8$\,\msun, straddling the radio jet downstream from MO at the location where the jet changes direction and decollimates. The clear evidence for interaction suggests that the \hi\ cooled out of the intergalactic medium (IGM) as a result of the jet's passage. 

\citet[][hereafter F17]{Fragile17} ran three-dimensional, multi-physics simulations, including a prescription to track star formation, of the collision of a jet with an intergalactic cloud, tailored specifically to match the measured parameters of MO, and found that indeed shocks triggered by the jet interactions with the cloud condense the gas and trigger cooling instabilities. Runaway increases in density result, and individual clumps become Jeans unstable. The resulting star formation rate, mass converted to \hi, molecular hydrogen, and stars, and the relative velocities of the stars and gas, are in good agreement with observations of MO, adding further weight to the JISF explanation for its properties.

In the present paper, we add another analysis tool to our study of MO. Radio polarimetry is a powerful technique for understanding the magneto-ionic medium associated with, and outside of, the synchrotron emitting electron population. Recent studies using broad-band or multi-band radio data are beginning to show the details of the magneto-ionic medium and its interaction with the IGM \citep{2008MNRAS.391..521L,2018ApJ...855...41A,2018MNRAS.476.1596K,2019MNRAS.482.5250B,2019A&A...622A.209A}. In this work, we report new VLA observations from 4--18\,GHz. When combined with archival data from lower and higher frequencies, we are able to construct a relatively complete picture of the radio brightness, spectral index, spectral curvature, and polarization. 

We combine these radio observations with new numerical simulations that capture the jet physics and corresponding radio emission self-consistently. By comparing simulations with and without a MO-like cloud, we are able to definitively demonstrate that the jet from NGC\,541 is currently interacting with MO.

The remainder of this paper is organized as follows: \S \ref{obs} describes the new VLA observations and their analysis. In \S \ref{numercialMethods} we outline the numerical simulations performed. In \S \ref{results} we present the results of our analysis of the simulations and compare their observable properties to those of NGC\,541 and MO. We summarize our main conclusions in \S \ref{conclusions}.

\section{Radio Observations} \label{obs}

\subsection{New VLA Observations} \label{VLAObs}

Minkowski's Object was observed in program 17A-109 on 2017-02-11 in Ku-band (12--18\,GHz) in D-configuration and on 2017-05-19 with the VLA in C-band (4--8\,GHz) in C-configuration. Time on source was 0.43\,hr in Ku-band and 2.15\,hr in C-band. 3C\,48 was used as a bandpass and flux calibrator, and J0125-0005 was used to calibrate the complex gains. For the Ku-band observations, the zero polarization calibrator  (3C\,84) was used to calibrate the polarization leakage. For the C-band observations, we took advantage of the longer scheduling block duration to observe the calibrator J\,0112+2244 over a range of 107 deg of parallactic angle to calibrate the polarization leakage. In both cases, 3C\,48 was used to calibrate the polarization position angle.\footnote{ \url{https://science.nrao.edu/facilities/vla/docs/manuals/obsguide/calibration}}\footnote{These data were taken prior to flaring activity in 3C~48, which began in January 2018.}

\subsection{Analysis}\label{radioAnalysis}

Calibration and imaging were performed in CASA (version 6.1.0-118) \citep{2007ASPC..376..127M}. Flux density, bandpass and complex gain calibration were performed for both datasets using the CASA 5.6.1 pipeline, with additional manual flagging to remove bad data and radio frequency interference. The pipeline currently does not calibrate the cross-hand delays or polarization leakage, so those were calibrated using the procedure described in the 3C\,75 polarization CASA Guide.\footnote{\url{https://casaguides.nrao.edu/index.php?title=CASA_Guides:Polarization_Calibration_based_on_CASA_pipeline_standard_reduction:_The_radio_galaxy_3C75-CASA5.6.2}} Multiband delays, averaged over each of the basebands (two for the C-band data and four in Ku-band), were derived to improve the signal-to-noise ratio of the cross-hand delays. 

In C-band, bright emission from the nearby radio galaxy 3C~40B (NGC\,547) is within the primary beam and needs to be accurately imaged so it can be correctly cleaned. We thus used $w$-term projection with 128 $w$-planes to correct for wide-field imaging effects. Multi-term, multi-frequency synthesis (MTMFS) imaging was used with three Taylor terms and imaging in Stokes I, Q, U and V to provide a model across the 4--8\,GHz bandwidth for self-calibration. Initially, three iterations of phase-only self-calibration were performed. Bandpass self-calibration was then performed, followed by a final iteration of amplitude and phase self-calibration. In Ku-band, the imaging was much simpler as NGC\,547 was outside the primary beam. No $w$-plane correction was needed and MTMFS imaging with two Taylor terms was sufficient. Three iterations of phase self-calibration and a single iteration of amplitude and phase self-calibration were used. All images were convolved to a common circular Gaussian beam of 6.5-arcsec FWHM for analysis.

\section{Numerical Methods} \label{numercialMethods}

We performed a series of numerical simulations using Wombat, the Eulerian, ideal, 3D, non-relativistic, magnetohydrodynamic (MHD) code described by \cite{Mendygral12}, originally based on \citet{Ryu95,Ryu98} and utilized, for example, in \citet{Jones17,Nolting19a}.
These simulations build upon the work of F17, but are distinct in their goals and included physics. In particular, they include magnetic fields in the jets as well as a population of cosmic ray electrons (CRe), which allow for the calculation of radio synchrotron emissivities throughout the simulated volume and the generation of synthetic radio observations. Because of this, we focus here on the radio observable properties of the simulation for comparison to the VLA observations described in the previous section.

The simulations were performed on a uniform, Cartesian grid using an adiabatic equation of state with $\gamma = 5/3$. The simulation utilized the 2nd-order TVD algorithm with constrained magnetic field transport as in \cite{Ryu98}. The simulation also incorporated a conservative Eulerian Fokker-Planck solver for transport of the CRe distribution, $f(p)$, employing the ``coarse grained momentum volume transport'' (CGMV) algorithm \citep{Jones99,JonesKang05}. The CGMV method evolves a CRe population over a wide range of particle momenta (or energies). We outline here only a few essentials of the method, referring readers to \cite{JonesKang05} and references therein for further details. In the current simulations, the isotropic CRe momentum distribution, $n_\mathrm{CRe} \propto \int p^2 f(p) \mathrm{d}p = \int p^3 f(p) \mathrm{d}(\ln p)$, was discretized within the range $30 \la p/(m_e c)\la 5.5\times 10^5$ (so, energies in the range 15\,MeV $\la E_\mathrm{CRe} \la$ 275\,GeV) with eight uniform, logarithmic momentum bins. Within a given bin, $k$, the momentum distribution is assumed to be a power law, $f(p) \propto p^{-q_k}$, though $q_k$ is allowed to vary between bins. 
The evolution of this distribution includes adiabatic and radiative (synchrotron and inverse Compton) energy exchanges outside of shocks, along with test-particle diffusive shock (re)acceleration (DSA) at shocks. The hydrodynamic shock detector finds the Mach number, $\mathcal{M} = v/c_s$, where $c_s$ is the local sound speed, based on the temperature jump across the shock following \cite{Skillman08}. Energy losses from Coulomb collisions are not included. For the thermal plasma densities relevant to these simulations, collisional losses are subdominant to inverse Compton/synchrotron losses unless $E_\mathrm{CRe} \la 200$\,MeV \citep[e.g.,][]{Sarazin99}, whereas the synchrotron emission examined in this study generally involve $E_\mathrm{CRe} \ga 1$\,GeV. For $E_\mathrm{CRe} \la 200$\,MeV, energy loss time scales generally exceed the simulation times in any case. For the simulated inverse Compton losses, we take the redshift of Minkowski's Object to be $z = 0.019$. For local magnetic field strengths less than $B \approx 3.25(1+z)^2\mu \mathrm{G} \approx 3.37 \mu \mathrm{G}$, inverse Compton losses dominate. For fields larger than this, synchrotron losses are more significant.

 In the presence of a shock, the test particle spectral index for DSA, $q_{k,\mathrm{out}} = \min\left[q_{k,\mathrm{in}},3\sigma /(\sigma - 1)\right]$, was applied immediately post-shock, where $\sigma$ is the code-evaluated compression ratio of the shock. This simple treatment is appropriate in the CRe energy range covered, since typical DSA times at those energies are much less than a typical time step in the simulation ($\Delta t \ga 10^3$ yr). The radio galaxy supposedly interacting with MO, NGC\,541, is an ``FRI'' radio galaxy, in which the relativistic plasma on multi-kpc scales is energetically subdominant \citep[e.g.][]{CrostonHardcastle14}.
 Accordingly, our CRe populations are passive, and the total CRe number density, $n_\mathrm{CRe}$, is arbitrary, since it has no impact on the dynamical evolution. Consequently, we can compute meaningful synchrotron brightness, polarization and spectral distributions from our simulations. To make our simulated radio maps most directly comparable to MO, we scale the radio intensities up by a factor of 4, corresponding to an increase in the density of CRe in our jets, which remain passive.

Except for a negligible background CRe population included in the ICM to avoid numerical singularity problems in the CGMV algorithm, all the CRe in our simulations were injected onto the computational domain as part of the AGN jet generation process (outlined in the following subsection). 
At injection from the AGN source, the CRe momentum distribution is assumed to be a power law with $q = q_0 = 4.2$, over the full momentum range. This translates into a synchrotron spectral index ($F_{\nu} \propto \nu^{-\alpha}$) of $\alpha = 0.6$, using the conventional relation for power laws. This was chosen to closely match the radio spectrum in the NGC\,541 jet near the host galaxy.

\subsection{Synthetic Radio Images}\label{syntheticObs}

The synthetic synchrotron emission we report are computed using the actual $f(p)$ over the momentum range specified above along with the full synchrotron emissivity kernel for isotropic electrons, given a local magnetic field $\vec{B}$. The synchrotron emissivity is \citep{Ginzburg65,Longair11}:

\begin{subequations}
 \begin{align}
j_\perp(\nu) & =  \sqrt{\frac{\nu \nu_{B\perp}}{8}} \frac{e^2}{c}  \int_{x_1}^{x_2} \frac{n(x) [F(x) + G(x)]}{x^{3/2}} \mathrm{d}x \label{eq:jperp} ~,\\
j_\parallel(\nu) &  =  \sqrt{\frac{\nu \nu_{B\perp}}{8}} \frac{e^2}{c}  \int_{x_1}^{x_2} \frac{n(x) [F(x) - G(x)]}{x^{3/2}} \mathrm{d}x \label{eq:jpar} ~,\\
j(\nu) & = j_\perp(\nu) + j_\parallel(\nu) =  \sqrt{\frac{\nu \nu_{B\perp}}{2}} \frac{e^2}{c}  \int_{x_1}^{x_2} \frac{n(x) F(x)}{x^{3/2}} \mathrm{d}x \label{eq:jnu} ~,\\
F(x) & = x \int_x^\infty K_{5/3}(z) \mathrm{d}z \approx 1.78 \times \left(\frac{x}{1 - 0.4 \exp{(-5x)}}\right)^{1/3}\exp{(-x)} \label{eq:Ffit} ~,\\
G(x) & = x K_{2/3}(x) \approx 1.56613 \times \frac{x^{1/3}}{\exp{(x) + 0.427687}} ~,\label{eq:Gfit}
 \end{align}
\end{subequations}
%%% commenting out the below version
\iffalse
\begin{subequations}
 \begin{align}
j_\perp(\nu) & =  \sqrt{\frac{\nu \nu_{B\perp}}{32}} \frac{e^2}{c}  \int_\infty^0 n(z) [F(z) + G(z)] \mathrm{d}z \label{eq:jperp} ~,\\
j_\parallel(\nu) &  =  \sqrt{\frac{\nu \nu_{B\perp}}{32}} \frac{e^2}{c}  \int_\infty^0 n(z) [F(z) - G(z)] \mathrm{d}z \label{eq:jpar} ~,\\
j(\nu) & = j_\perp(\nu) + j_\parallel(\nu) =  \sqrt{\frac{\nu \nu_{B\perp}}{8}} \frac{e^2}{c}  \int_\infty^0 n(z) F(z) \mathrm{d}z \label{eq:jnu} ~,\\
F(x) & = x \int_x^\infty K_{5/3}(z) \mathrm{d}z \approx 1.78 \times \left(\frac{x}{1 - 0.4 \exp{(-5x)}}\right)^{1/3}\exp{(-x)} \label{eq:Ffit} ~,\\
G(x) & = x K_{2/3}(x) \approx 1.56613 \times \frac{x^{1/3}}{\exp{(x) + 0.427687}} ~,\label{eq:Gfit}
 \end{align}
\end{subequations}
\fi
%%
where $F(x)$ and $G(x)$ are functions that describe the synchrotron spectrum of a single CRe, $K_{5/3}$ and $K_{2/3}$ are modified Bessel functions of the second kind, $j_\perp(\nu)$ and $j_\parallel(\nu)$ are the two orthogonal polarizations of the emissivity, $\nu$ is the frequency of the radiation, $\nu_{B\perp} = eB_\perp/2\pi m_e c$ is the electron gyrofrequency, and $e$ is the elementary charge. In this case, $B_\perp$ is the local magnetic field projected onto the plane of the sky.
The integration variable $x = (2\nu/3\nu_{B\perp}\gamma^2)$, with $\gamma$ being the electron Lorentz factor. The integration limit $x_1$[$x_2$] corresponds to the high[low] energy (high[low] $\gamma$) end of the integration range, covering our available range $30 < p/(m_ec) < 5.5 \times 10^5$. A change of variables from integrating over $\gamma$ to integrating over $x$ introduces the factor of $x^{-3/2}$ in the integrand and a coefficient of $(-1/2)\sqrt{2\nu/3\nu_{B\perp}}$. We also present these equations in Gaussian units, rather than the SI units used by \cite{Longair11}.
Equations (\ref{eq:Ffit}) and (\ref{eq:Gfit}) include our own approximations to the synchrotron functions used in our calculations. These approximations are based on those given in \cite{RybickiLightman86}, but combine the small $x$ and large $x$ approximations to produce a single fit that is accurate to a few percent \citep{NoltingThesis}. 

Using equations (\ref{eq:jperp}) - (\ref{eq:Gfit}) we calculate the total synchrotron emissivity as well as polarized emissivities. Then, we perform radiative transfer integration along a defined line of sight to create images of Stokes I, Q, or U. 

A radio spectral index is calculated from any two radio maps at different frequencies with the radio spectral index, $\alpha$ (e.g., $f\propto \nu^\alpha$), at each pixel being
\begin{equation}\label{eq:index}
    \alpha = \frac{\log_{10}(I(\nu_2) - \log_{10}(I(\nu_1)))}{\log_{10}(\nu_2) - \log_{10}(\nu_1)}.
\end{equation}
The radio spectral curvature, $\beta$, is calculated by fitting the intensity at any three frequencies to a function:
\begin{equation}\label{eq:curvature}
    \log_{10}(f(\nu)) = \log_{10}(f(\nu_0)) + \alpha \log_{10}(f(\nu)) + \beta \log_{10}^2(f(\nu)),
\end{equation}
where $f(\nu_0)$ serves as a normalization for the functional fit and $\alpha$ is again the radio spectral index, here left as a fitting parameter alongside the curvature, $\beta$.

Synthetic images were calculated assuming a redshift equal to that of MO (z = 0.019), which corresponds to an angular diameter distance of 83\,Mpc.\footnote{Here we assume a cosmology with $H_0$ = 67.74 (km/s)/Mpc \citep{Planck2015}.} This meant that an image pixel, which represented a projected physical size of (78.1/156.3\,pc) depending on the presented simulation (see Table \ref{tab:simtable}), subtended an angular width of (0.2/0.4) arcseconds. To better compare with observations, we convolve these images with a Gaussian smoothing kernel to simulate lower angular resolution\footnote{Gaussian2DKernel from AstroPy.Convolution \citep{2013A&A...558A..33A}}. 

\subsection{Simulation Setup}\label{setup}

We present four 3D simulations of jet-cloud interactions. Our study is based on the earlier work of F17, but tries to converge to a more realistic dynamical scenario. Similar to F17,
a dense cloud with a radius of $R_\mathrm{cl} = 7.5$\,kpc is initialized on the grid. The gas in the cloud itself is non-gravitating, but is initialized in hydrostatic equilibrium with a static dark-matter potential associated with the cloud (not the cluster potential). The gravitational potential is given by a modified Hubble profile:
%%%%
\begin{equation}
    \phi(r<R_t) = \frac{G\tilde{M}}{R_c}\bigg\{1 - \frac{\ln\left[x+(1+x^2)^{1/2}\right]}{x}\bigg\},
\end{equation}
%%%%
where $x=r/R_c$, $R_c = 0.5R_\mathrm{cl}$ is the core radius, and 
%%%%
\begin{equation}
    \tilde{M} = M_d \bigg\{ \ln\left[x_t + (1+x_t^2)^{1/2} \right] - x_t(1+x_t^2)^{-1/2} \bigg\}^{-1},
\end{equation}
%%%%
where $M_d = 10^{11}M_\odot$ is the dark-matter mass, $x_t = R_t/R_c$, and $R_t = 10R_\mathrm{cl}$ is the tidal radius. The cloud is initialized to be isothermal, with $T_\mathrm{cl} = 2\times10^5$\,K, and in hydrostatic equilibrium with the potential such that the density
%%%%
\begin{equation}
    \rho_\mathrm{cl} \propto e^{-\phi/c_s^2},
\end{equation}
%%%%
where $c_s$ is the isothermal sound speed. Differing from F17, we do not include clumping within the cloud, as we are not focusing here on star formation and do not include the effects of cooling inside the cloud. The cloud mass is normalized such that the total gas mass within $R_\mathrm{cl}$ is $M_g = 1.4\times 10^9 M_\odot$, and gives the cloud an average density of $\bar{\rho}_\mathrm{cl} = 5.3\times 10^{-26}$g cm$^{-3}$. The cloud is surrounded by a uniform background gas with number density $n_b = 10^{-4}$ cm$^{-3}$ and temperature $T_b = 5\times10^7$\,K, so that at the cloud surface the pressure is in approximate equilibrium. Both the cloud and the surrounding medium are initially unmagnetized.

The four simulations presented here have different properties which are briefly summarized in Table \ref{tab:simtable}. The first new simulation follows the setup of F17 closely, with a stationary cloud and straight jet. Simulations without significant bending of the jets are labeled simply with ``\textbf{JET}'' while simulated jets that are bent by a strong wind into narrow angle tail sources are labeled with ``\textbf{BENT}.'' If a dense cloud is present in the simulation, the simulation name will also have ``\textbf{SC},'' if the cloud is stationary, or ``\textbf{MC},'' if it is in motion.

\begin{table}[]
    \centering
    \begin{tabular}{|c|c|c|c|c|c|c|}
    \hline
    Simulation & Box Size & Resolution & Jet & Jet Mach & Wind & Cloud Velocity\\
     Name & (kpc) & (pc)  & Velocity & Number & Velocity & (km/s)\\
    \hline
    \textbf{JETSC} & $37.5 \times 31.25 \times 31.25$ & 78.1 & 0.1 c & 8.2 & 0 km/s & $v_x = 0$, $v_y = 0$\\
    \textbf{JETMC} & $78.75 \times 101.25 \times 48.75$ & 156.3 & 0.1 c & 8.2 & 250 km/s & $v_x = 0$, $v_y = 250$\\
    \textbf{BENTMC} & $146.25 \times 140 \times 48.75$ & 156.3 & 0.042 c & 3.4 & 900 km/s & $v_x = 500 $, $v_y = -250 $ \\
    \textbf{BENT} & $146.25 \times 140 \times 48.75$ & 156.3 & 0.042 c & 3.4 &900 km/s & N/A\\
    \hline
\end{tabular}
    \caption{Key parameters of the four simulations. The jet Mach number, $\mathcal{M}_j$, is with respect to the internal sound speed $c_{s,j} = \sqrt{\gamma P_j/\rho_j}$. Cloud $v_x$ and $v_y$ refer to the horizontal and vertical velocities, respectively, as shown in Figure \ref{fig:mvhr2-spec} and the top two panels of Figure \ref{fig:specIndexCombined}}.
    \label{tab:simtable}
\end{table}

In all four simulations, a bilateral jet is injected into the simulation domain. The jet is launched from a cylindrical region within the domain. For \textbf{JETMC}, \textbf{BENT}, and \textbf{BENTMC} the jet launch cylinder was 28 grid cells long with a radius of 16 grid cells, representing a jet diameter of 5\,kpc. \textbf{JETSC} had twice this resolution across the jet. The cylinder was surrounded by a coaxial cylindrical collar two grid cells thick providing a transition region from the conditions maintained inside the jet launch cylinder and the ambient conditions in the surrounding medium. The jets consisted of a non-relativistic (thermal), weakly magnetized [$\beta_{j} = P_j/(B_j^2/8\pi)$ = 75] plasma with a density of $\rho_j = 1.1\times 10^{-29}$g cm$^{-3}$, making it about $20 \times$ less dense than the ambient ICM. The jet was also initially in pressure equilibrium with the ICM, and launched with a velocity of $ v_j = 0.042 c = 1.25\times 10^9 $ cm s$^{-1}$  for the \textbf{BENTMC} and \textbf{BENT} simulations and $v_j = 0.1 c = 3\times 10^9$ cm s$^{-1}$ for the \textbf{JETSC} and \textbf{JETMC} simulations (see Table \ref{tab:simtable}). This gives the jet an internal Mach number $\mathcal{M}_j = v_j/\sqrt{\gamma P_j / \rho_j} = 3.4$ or $8.2$, respectively. Each jet also contained a passive tracer quantity and a population of passive, relativistic CRe in order to model non-thermal emission. The passive tracer represents the ``jet mass fraction,'' as it is set to unity for any material entering the grid via the jet cylinder, but initialized to zero everywhere else. The jet magnetic fields at launch are toroidal, created by a uniform poloidal electric current aligned with the jet velocity. A poloidal return current exists in the launch cylinder transition collar, so that the net electric current along the jet launch cylinder vanishes everywhere. With the plasma beta $\beta_{j}$ = 75, the jet field strength at the source is $B_j = 0.54 \mu$G.

To bend the jets into a narrow angle tail source, for the \textbf{BENT} and \textbf{BENTMC} simulations the medium was also put into motion. This relative motion between the medium and the jet source represents a combination of the AGN's host galaxy's orbit and the bulk motion of the ICM in Abell 194 near MO. This ``wind'' is initialized perpendicular to the jet axis and creates ram pressure that bends the jets until they are nearly parallel to the wind propagation direction. This bending happens on a scale, $l_b$, defined by the ratio of the momentum fluxes in the jet and the wind \citep{Begelman79,Oneill19}:
\begin{equation} \label{eq:bending}
l_b = r_j \frac{\rho_j v_j^2}{\rho_w v_w^2} ~,
\end{equation}
where $r_j$ is the radius of the jet, $\rho_w$ is the density of the wind, and $v_w$ is the velocity of the wind. Given the density and velocity of the ICM material, the bending scale for the jets in \textbf{BENT} and \textbf{BENTMC} is $l_b = 9.8 r_j$.

In \textbf{JETMC} and \textbf{BENTMC}, the dense cloud is put in motion on a collision course with the jet to begin the jet-cloud interaction. The cloud initial velocity is set to the values listed in Table \ref{tab:simtable}. The (fixed) gravitational potential is advected across the grid with the same velocity as the initial cloud velocity. The cloud in the \textbf{JETMC} simulation was also offset along the line of sight in Figure \ref{fig:mvhr2-spec} (simulation z-axis) from the center of the jet by a distance equal to half the cloud radius. This made for more of a ``glancing'' style collision rather than the jet impacting directly onto the core of the cloud, as is done in \textbf{JETSC}. Lastly, the \textbf{JETMC} simulation had a light wind designed not to bend the jet, but to blow away the cocoon of radio plasma from previous jet activity that would otherwise build up surrounding the jet. No cocoon is present surrounding the jet in NGC\,541, so this wind helped us match the observed radio morphology.

\section{Results} \label{results}

\subsection{Early Simulation Attempts}

\begin{figure}
    \centering
    \includegraphics[width=0.7\textwidth]{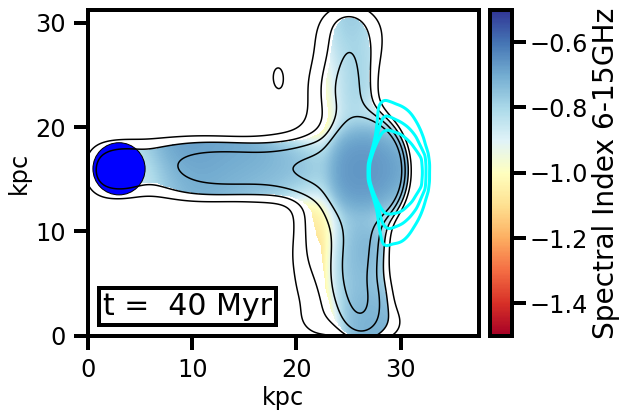}
    \caption{Simulation \textbf{JETSC} at $t = 40$\,Myr from the start. The radio spectral index (6--15\,GHz, cut at 10$\mu Jy/\text{beam}$ at 15GHz) is shown, along with 6 GHz radio brightness contours (levels = [40, 80, 160, 320]$\mu Jy/\text{beam}$)  in black and projected surface density contours (levels = [5, 10, 20]$\times 10^{-4}$ g/cm$^2$) to represent the dense cloud in cyan. A blue circle marks the location of the jet source. All radio quantities are convolved to 6.5'' resolution.}
    \label{fig:jetcl}
\end{figure}

Here we will describe the results of the simulations and compare them with the observations of Minkowski's Object and the radio jet from NGC\,541. To start, the \textbf{JETSC} simulation was an early attempt to model the radio emission from a head-on collision between a jet and a cloud akin to the simulations of F17. Figure \ref{fig:jetcl} shows the radio spectral index between 6\,GHz and 15\,GHz as well as 6\,GHz radio intensity contours and contours of projected cloud surface density from the simulation. These surface density contours show the location of the cooler, denser gas and serve as a tracer for HI later when we compare to MO. In the \textbf{JETSC} simulation, the radio jet source is on the left and propagates to the right and interacts with the cloud. We show the properties of the simulation 40\,Myr after it started, which is about 35\,Myr after the jet-cloud collision began. There were a number of issues with making comparisons between this simulation and the jet from NGC\,541. First, in the case of NGC\,541's jet, the radio emission clearly extends past MO, as can be seen in the bottom panel of Figure \ref{fig:specIndexCombined}. However, when the simulated jet impacts directly onto the core of the dense cloud, it is stopped by the strong pressure gradient in the cloud and deflected laterally, such that much of the jet plasma then flows directly off of the simulation grid. This also causes the outer layers of the cloud to be deformed by the interaction and spread out laterally. Furthermore, at the interaction point between the jet and the cloud, there is a shock ($\mathcal{M} \approx 6$) that compresses the jet plasma and amplifies the fields. This leads to a radio hotspot that is about 2 orders of magnitude brighter than the surface brightness in the jet itself. Additionally, since the jet plasma flows quickly away from the interaction point and off the grid, there is not sufficient time for the radio plasma to cool and age, so it retains a flat spectrum without any significant steepening. None of this is consistent with what is seen in MO. 

In order to avoid these issues in subsequent simulations, we tried introducing a physical offset between the core of the cloud and the jet so the interaction was more ``glancing.'' We ran a number of these with stationary clouds, but none were able to match the morphology of MO, particularly the emission beyond MO and the radiatively aged material as evidenced by the spectral index. This led us to try starting the cloud outside the path of the jet, but in motion towards it, so that the jet had time to establish itself before the cloud interaction began and the interaction could begin more gradually. We report on only one of these attempts -- the simulation \textbf{JETMC}. In this simulation, we initialized the cloud outside the path of the jet and set it in motion towards it. The hope was this would allow the jet to fully develop before the interaction occurred, which would allow some jet material to age and a portion of the jet to continue past the dense cloud.
%, as the radio jet in NGC\,541 continues beyond MO. 
This would also allow us to capture different stages of a slow, progressing encounter to see if any stage matched well with the morphology of MO. The motion of the cloud is also more physically motivated by the fact that in any galaxy cluster environment, any dense cloud of material is likely to have some orbital motion on the order of a few tenths of the sound speed in the medium. The path of the cloud in \textbf{JETMC} was also offset from the axis of the jet by $3.25$\,kpc, or half a cloud radius. This was done to decrease the severity of the interaction and thus decrease the brightness of the radio hotspot at the interaction site. Lastly, a light wind in the medium was added with velocity $v_w = 250$\,km/s. This did not have enough ram pressure to significantly alter the trajectory of the jet, but did act to displace the radio cocoon that would otherwise build up around the jet. 

\begin{figure}
    \centering
    \includegraphics[width=0.7\textwidth]{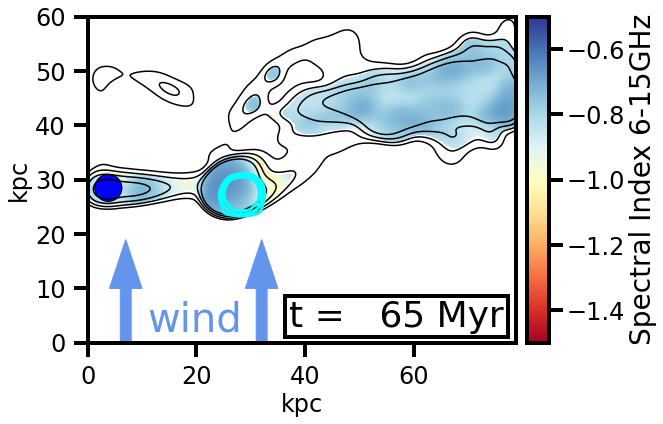}
    \caption{Simulation \textbf{JETMC} at $t = 65$\,Myr from the start of the simulation. The radio spectral index (6--15\,GHz, cut at 10$\mu Jy/\text{beam}$ at 15GHz) is shown, along with 6 GHz radio brightness contours (levels = [40, 80, 160, 320]$\mu Jy/\text{beam}$) in black and projected surface density contours (levels = [5, 10, 20]$\times 10^{-4}$ g/cm$^2$) to represent the dense cloud in cyan. A weak wind in the vertical direction (simulation y-axis) inflows from the lower boundary to displace the initial radio cocoon plasma without significantly bending the jet. A blue circle marks the location of the jet source, at the left of the figure pointed toward the right. All radio quantities are convolved to 6.5'' resolution. In the online version, an animated figure (10 sec) displaying the evolution of the simulation from $t=0$ to 80 Myr is available. The cloud begins outside the jet and moves at the wind speed into the path of the jet. The jet interaction begins around 45 My, at which point the jet begins to deflect towards the upper right of the image. The cloud has fully entered into the path of the jet by 70 Myr,  which causes the jet material to rebound laterally.  \url{https://www.youtube.com/watch?v=8zTEUFpH1rU}}
    \label{fig:mvhr2-spec}
\end{figure}

These changes led to some notable improvements in the radio properties of the \textbf{JETMC} simulations as shown in Figure \ref{fig:mvhr2-spec}. At the time presented ($t = 65$\,Myr from the start of the simulation, which is about $25$\,Myr since the cloud first started interacting with the jet), the dense cloud has not yet made its closest approach to the core of the jet. The radio plasma is seen rebounding off the outer portions of the cloud. Since the cloud is offset toward the observer by half a cloud radius, the rebounding jet is angled upward and into the page in the image.
At the interaction site, a shock ($\mathcal{M} \approx 6.2$) reaccelerates the CRe, flattening the spectrum, although just behind the cloud there is some steep spectrum emission visible from aged jet plasma along the line of sight.
The jet plasma rebounding away from the interaction site expands, lowering the magnitude of the magnetic fields, thus lowering the synchrotron emissivity, leading to a significant decrease in the brightness of the jet material. Ultimately the rebounding jet over-expands and subsequently collapses back in on itself, forming another shock ($\mathcal{M} \approx 6.7$) where the fields are amplified again to values comparable to those in the original jet. This again leads to a redistribution of the CRe due to DSA to a spectral index of $\alpha \approx 0.52$. This reacceleration, as well as the higher field values, produce another region of radio emission downstream of the cloud with a radio spectrum that is too flat when compared with what is seen in MO (see in the upper right of Figure \ref{fig:mvhr2-spec}). Furthermore, although the radio hot spot where the jet meets the cloud is reduced considerably relative to the \textbf{JETSC} simulation, it is still too bright to properly represent the interaction with MO.

\subsection{Bent Jet Simulations}\label{bentJets}

\begin{figure}
    \centering
    \includegraphics[width=0.7\textwidth]{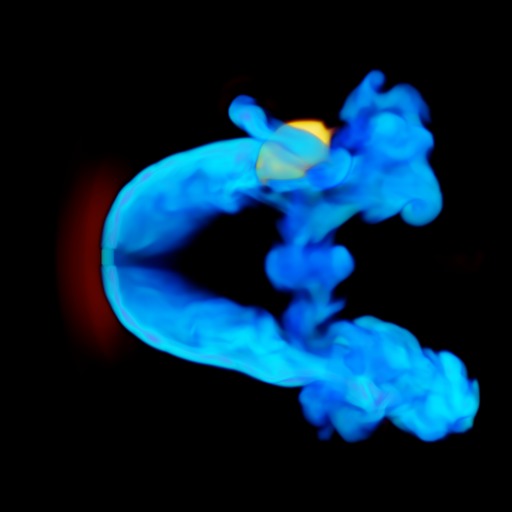}
    \caption{ 3D rendering of the logarithmic denisty of the \textbf{BENTMC} simulation at t=80Myr. The density of the medium is masked out to highlight the low density jets(blue) and the high density cloud (orange) representing MO. The cloud is initialized in the upper left corner and in motion down and to the right to interact with the jets. The jets are launched vertically and are bent into tails by a strong horizontal wind. An animated version of this figure (10 sec) displaying the evolution of the simulation from $t=0$ to 80 Myr is available online. \url{https://www.youtube.com/watch?v=pwq-xK9dnP0}}
    \label{fig:bentmc-logrho-3d}
\end{figure}

Looking more closely at the actual morphology of the jet in NGC\,541 (bottom panel of Figure \ref{fig:specIndexCombined}), there is strong evidence we are seeing a bent tailed jet source in projection. It could be that this bending of the jet plays an important role in determining the nature of its interaction with MO. To test this idea, we added a wind (stronger than that in \textbf{JETMC}) to the medium to bend our jet into a narrow angle tail (NAT) source. The wind was set to $v_w = 900$ km/s, such that it is still slightly subsonic. Highly bent radio galaxies can experience such strong winds (relative motion), especially if they are on radial infall orbits \citep{Pimbblet06,Garon19}. In particular, estimates of the dynamics of NGC 541's motion through Abell 194 give a rough estimate of its velocity of $v = 788^{+175}_{-438}$ km s$^{-1}$ \citep{Bogdan11}, compatible with our proposed wind speed. To reach the desired bending length, the ram pressure in the jet also had to be reduced. Since the numerical time step is set based on the sound speed in the jet, we decided to fix the jet density to avoid increasing the computational costs of the simulations. This meant we needed to reduce the jet speed to $v_j = 0.042c$ so that the bending length, given by Equation (\ref{eq:bending}), is 32 kpc, which is approximately equal to our best guess for the de-projected bending length of the NGC\,541 jets. 
We also rotated the observer in our synthetic images to match the apparent projected viewing angle of NGC\,541.
To clearly establish the effect of the cloud on the jet morphology in this case, we ran two different simulations, one with a cloud (\textbf{BENTMC}) and one without (\textbf{BENT}). 
Figure \ref{fig:bentmc-logrho-3d} shows a 3-dimensional rendering of the logarithmic density in the \textbf{BENTMC} simulation at t=80Myr. In \textbf{BENTMC}, the dense cloud (shown as cyan logarithmic density contours) is initialized in what would be the upper left hand corner of the top panels in Figure \ref{fig:specIndexCombined} and moves with a velocity that carries it along with the wind but also into the path of the bent jet. The rest of this section presents detailed comparisons between the synthetic radio properties of the \textbf{BENTMC} and \textbf{BENT} simulations and the VLA observations of MO (as described in \S \ref{obs}).

\subsubsection{Spectral Index}\label{index}

Overall, there is good morphological agreement between the NGC\,541 jet and the \textbf{BENTMC} simulation. The eastern jet propagates toward the dense cloud (MO) and then deflects near the position of the cloud. The western jet bends backward and is seen in projection behind and north of the eastern jet. One striking difference between the observations and simulations is in the surface brightness near the core of the jet. In the simulated images, for both \textbf{BENTMC} and \textbf{BENT}, when rotated to the projected view, the jet appears strongly edge brightened. This is due to the alignment of the magnetic fields in the jet after being bent backwards. The simulated jets are launched with purely toroidal field geometries, with the fields strongest at the jet edge and disappearing at the center. This is done mostly for computational convenience; this geometry is the simplest one to initialize that guarantees closed field lines separate from fields in the medium. However, as the jets bend, the asymmetric ram pressure from the wind causes the fields to be amplified strongly at the edge of the jet in contact with the wind due to compression and stretching in a shear layer. When viewed from this projected angle, there is a longer path length through these high field regions at the projected top and bottom edges of the jet than through the projected center of the jet, which is dominated by the lower magnitude fields near the physical core of the jet. The more uniform brightness in the NGC\,541 jet may suggest more uniform compression inside the jet during bending or some other field geometry that does not experience this effect.

\begin{figure}
\centering
    \includegraphics[width=\textwidth]{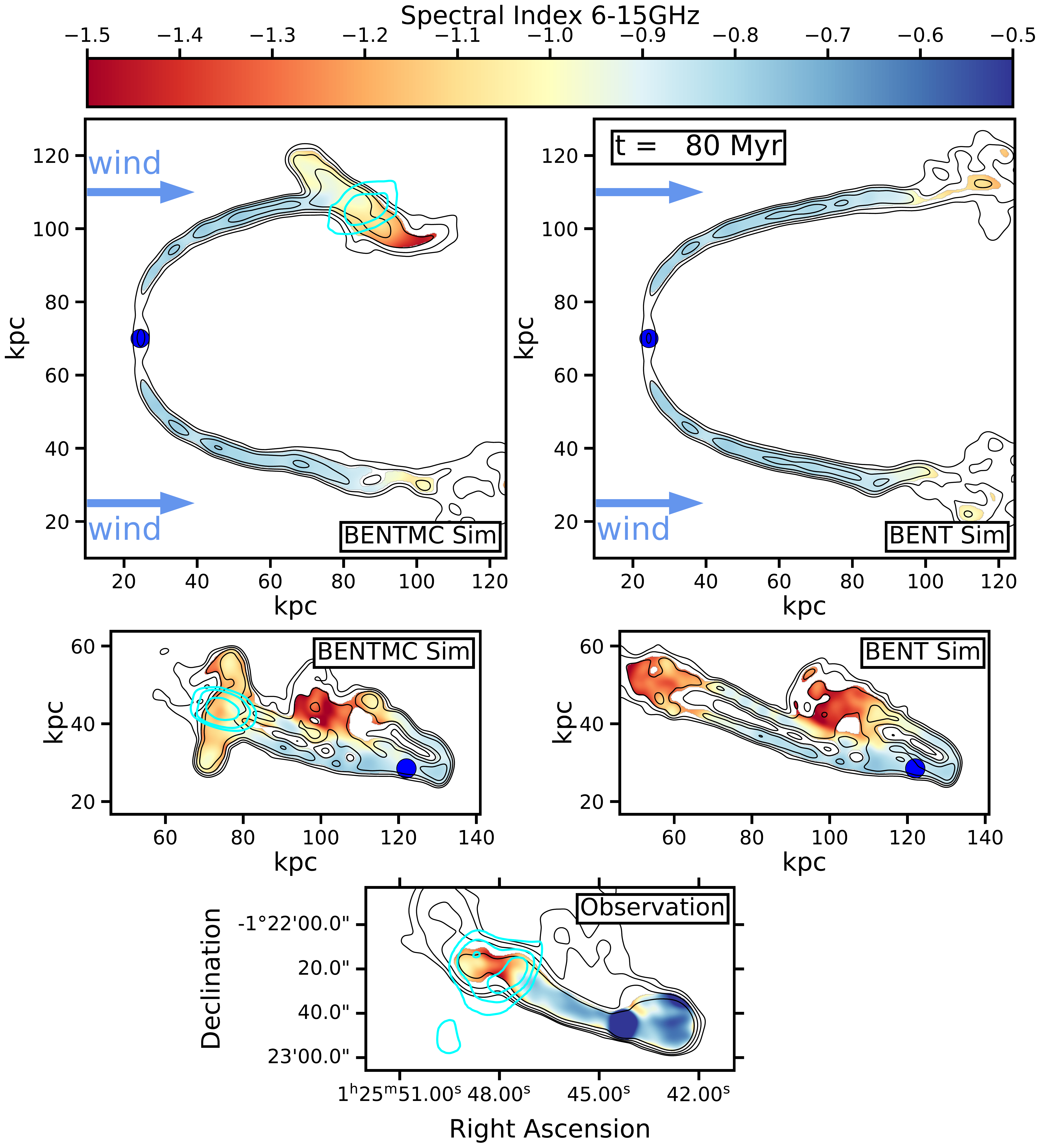}
    \caption{Radio spectral index (6--15\,GHz, cut at 150$\mu Jy/\text{beam}$ at 15GHz) maps with 6\,GHz radio brightness contours (levels = [150, 300, 600, 1200]$\mu Jy/\text{beam}$ in black) for the \textbf{BENTMC} (top- and middle-left panels) and \textbf{BENT} (top- and middle-right panels) simulations and for the observations of NGC\,541 (bottom panel). The cyan contours represent projected surface density of $T \sim 10^5$ gas (\textbf{BENTMC} simulation, levels = [5, 10, 20]$\times 10^{-4}$ g/cm$^2$) or surface density of \hi (MO, levels = [0.05, 0.075, 0.1, 0.125] $\text{Jy km/s/beam}$, \cite{croft:06}). In the top panels, a strong wind in the horizontal direction inflows from the left boundary to bend the simulated jets. In the middle panels, the simulated jets are rotated and projected to the suspected orientation of NGC\,541 with the winds mostly pointing into the page, up, and to the left. In the simulated images, the blue circle marks the location of the jet source. All radio quantities (simulated and observations) are presented at 6.5'' diameter beam size. The bottom panel cutout size is 166''x83.'' An animated version of this figure (10 sec) displaying the evolution of the simulations from $t=0$ to 80 Myr is available online. In both simulations, the jets are shown being bent backwards by the wind into tails. In the \textbf{BENTMC} simulation, the cloud comes in from the upper left and begins to interact with the jet around 45 Myr. The cloud has fully entered into the path of the jet by 65 My, causing the jet material to rebound laterally. \url{https://www.youtube.com/watch?v=4MEUi2W4v1A}}
    \label{fig:specIndexCombined}
\end{figure}

In Figure \ref{fig:specIndexCombined} we also observe that the simulated radio jet in \textbf{BENTMC} gets spectrally steeper at the cloud interaction site similar to what is seen for the NGC\,541 jet in the vicinity of MO. 
In the simulation, this is due to the jet's forward propagation being disrupted, causing older jet material to pile up near the cloud interaction point, as can be seen in the top left panel. In the \textbf{BENT} simulation, this material instead continues on and ages farther away from the jet source.
In the \textbf{BENTMC} simulation, since the velocity of the jet is reduced, and the cloud has a velocity component parallel to the wind, the interaction and deflection are less extreme.
This leads to a shock of reduced strength at the interaction point ($\mathcal{M} \approx 2.6$) and an increase in the radio brightness at that point by only about $\sim 2\times$ compared to the surface brightness in the jet prior to impact. DSA would predict a radio spectral index of $\alpha \approx 0.66$ for emitting plasma passing through this shock, which is not much flatter than the emission from the incoming jet. Because of this, we don't see significant flattening in this region from DSA, even though it does occur there in the simulations. Similarly, there is no spectral flattening observed beyond the interaction point, in contrast to the \textbf{JETMC} simulation, as the rebounding material does not collapse and form a secondary shock in this simulation. 

This behavior in the \textbf{BENTMC} simulation differs from the observations, in which the spectrum flattens from $\alpha \approx -1.3$ near the interaction site to $\alpha \approx -1$ in the deflection region. This steepening of the radio spectrum at the interaction site and flattening beyond it is difficult to explain if there is a shock at the interaction point that reaccelerates and redistributes the CRe energy distribution into a pure power-law, as expected by DSA. The flat region would be expected near the shock and it would steepen away from that region as radiative aging progresses.

 There are a number of possible explanations for this difference in the observations. First, in our simulations we explored only a limited portion of the parameter space for the properties of the jet, the wind, and the dense cloud. Different choices for the relative momentum of those features would lead to different (and possibly stronger or weaker) interactions -- including shock strength and the importance of DSA. If the underlying CRe distribution is not as efficiently reaccelerated by DSA in the NGC 541 jet as it is in our simulations, and instead the distribution remains curved due to radiative losses prior to the interaction, then the observed changes in spectal index might be explained by changes in the magnetic field strength. Cosmic rays emitting in higher field regions will emit predominantly higher frequencies, so that at a fixed observation frequency when we look at higher magnetic field regions we are looking at emission from lower energy electrons, which should have experienced less cooling from inverse Compton and synchrotron losses and result in a flatter spectrum. Thus, if there were higher fields at the interaction point due to compression and lower fields in the deflected (possibly expanding) jet, it could explain the spectral behavior in the observed jet. However, this explanation fails if the energy distribution is not curved and is instead a power-law, as expected if DSA occurs at the interaction site if a shock is present (as in the \textbf{BENTMC} simulation).

%To explain the observation, we consider a case in which the underlying CRe distribution is not efficiently reacccelerated by DSA and remains curved due to radiative losses prior to the interaction. In this case, the changes in spectal index might be explained by changes in the magnetic field strength. Cosmic rays emitting in higher field regions will emit predominantly higher frequencies, so that at a fixed observation frequency when we look at higher magnetic field regions we are looking at emission from lower energy electrons, which should have experienced less cooling from inverse Compton and synchrotron losses and result in a flatter spectrum. Thus, if there were higher fields at the interaction point due to compression and lower fields in the deflected (possibly expanding) jet, it could explain the spectral behavior in the observed jet. However, this explanation fails if the energy distribution is not curved and is instead a power-law, as expected if DSA occurs at the interaction site if a shock is present (as in the \textbf{BENTMC} simulation).

An alternative explanation is that our simulations fail to capture all emission mechanisms, including emission from newly formed stars or from supernovae from stars formed in the early stages of the encounter. Such emission may also explain the flattening observed in the deflection regions past MO seen in Figure \ref{fig:specIndexCombined}.

\subsubsection{Polarization}\label{polarization}

\begin{figure}
    \centering
    \includegraphics[width=0.7\textwidth]{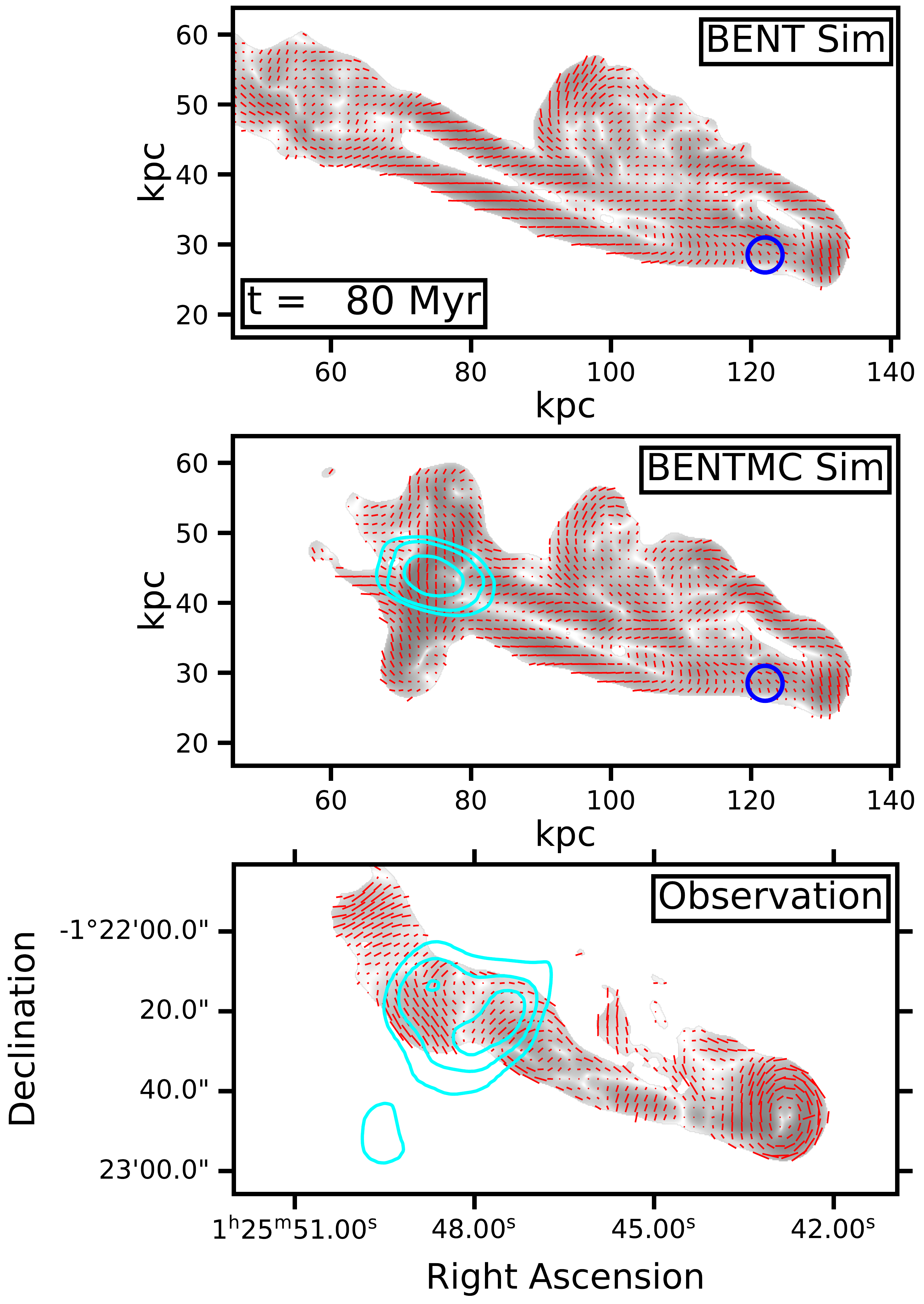}
    \caption{Logarithmic radio polarized intensity maps (6 GHz, cut at 150$\mu Jy/\text{beam}$) for the \textbf{BENT} (top) and \textbf{BENTMC} (middle) simulations and for the observations of NGC\,541 (bottom) shown in grayscale between 150 and 1000 $\mu Jy /\text{beam}$. Cyan contours are as in Figure \ref{fig:specIndexCombined}. Radio polarization vectors (rotated 90$^{\circ}$ to represent the magnetic field orientation) are shown with lengths proportional to the fractional polarization (above the same radio polarized intensity cutoff). In the simulated images, the blue circle marks the location of the jet source. The beam sizes in each panel and cutout size in the bottom panel are the same as in Figure \ref{fig:specIndexCombined}. An animated version of this figure (10 sec) displaying the evolution of the simulations from $t=0$ to 80 Myr is available online. The polarization in the \textbf{BENTMC} simulation begins to deviate from what is seen in \textbf{BENT} once the cloud starts interacting with the jet (at $t\approx 45$ Myr),  at which point the polarization vectors are roughly perpendicular to their previous orientation as the jet material rebounds laterally away from the interaction site. \url{https://www.youtube.com/watch?v=KXskWvp6j6U}}
    \label{fig:polarizationCombined}
\end{figure}

Figure \ref{fig:polarizationCombined} shows the radio polarization properties of NGC\,541 and the \textbf{BENTMC} and \textbf{BENT} simulations. In all three cases, the polarization vectors in the core of the jet near the jet source are perpendicular to the jet trajectory, implying a predominantly toroidal field configuration. In NGC\,541, the edges of the jet show polarization angles that are rotated 90$^\circ$ from this, which may imply there is stretching of the fields along the jet direction due to a shear layer between the core of the jet and the medium. This shear is present along the boundary of the jets in the simulation, and a poloidal component develops during the jet bending along the jet-medium shear boundary, which can be seen as a change in polarization angle towards the outer edges of the simulated jets. Recall the simulations show edge brightening due to field compression and shear in these same regions. Of much greater interest are the changes due to the presence of MO. The polarization angle in the NGC\,541 jet changes abruptly and dramatically near the core of MO. The jet material also changes its trajectory slightly at the same location. Beyond MO, the jet trajectory and polarization angle return to their pre-MO state. This behavior again matches well with the \textbf{BENTMC} simulation, as that simulation shows a clear change in the polarization angle at the location of the cloud interaction point, with some fairly well collimated jet material deflecting away from the interaction point toward the north during the early phases of the interaction (before the cloud core enters into the center of the jet). As such, the fields in the deflected jet remain predominantly toroidal, and the polarization vectors remain perpendicular to the propagation direction. Fields have become disordered in the farthest reaches of the non-interacting jets, in both \textbf{BENT} and \textbf{BENTMC}, due to instabilities in the trajectory of the end of the jet (so-called ``jet flapping,'' see \citealt{Oneill19}) that have transformed the well collimated jets into turbulent tails. 

A strong polarization feature also exists in the western jet of NGC\,541. Both the polarized intensity and fractional polarization are high near the location where the jet is being bent back strongly and the polarization vectors clearly encircle the core of the jet. This is strong evidence that the field configuration is toroidal and that we are ``looking down'' the jet propagation direction at that location before the jet bends away from our line of sight. The simulated images do not show a complete field circulation at the same dynamical point.
This may suggest that our jets in the \textbf{BENTMC} and \textbf{BENT} simulations are still not oriented quite the same relative to the observer as the jets in NGC\,541. However, the polarization vectors in the simulations at this location are oriented vertically and may correspond to the magnetic fields on the ``windward'' side\footnote{By windward side, we mean the portion of the structure closest to the left-hand side of the top panels of Figure \ref{fig:specIndexCombined}, i.e. the upwind direction.} of such a toroidal field structure. Since those fields are amplified significantly in the simulations by the ram pressure and shear from the medium, they dominate enough that contributions from other field orientations are not visible in our synthetic polarization vector images. 
 
Finally, an interesting feature shows up in the fractional polarization and polarized intensity (vector lengths and grayscale in Figure \ref{fig:polarizationCombined}, respectively). An ``x-shape'' feature appears in the eastern jet of NGC\,541 about 8\,kpc from the radio core. This is reminiscent of a recollimation shock in a jet. There are definitely recollimation shocks present in the simulated jets. However, they show up as radio brightenings without significant changes in the fractional polarization. Differences in the magnetic field configurations between the simulations and the jet of NGC\,541 may allow for this structure to still represent a recollimation shock.

\subsubsection{Spectral Curvature}\label{curve}

\begin{figure}
    \centering
    \includegraphics[width=\textwidth]{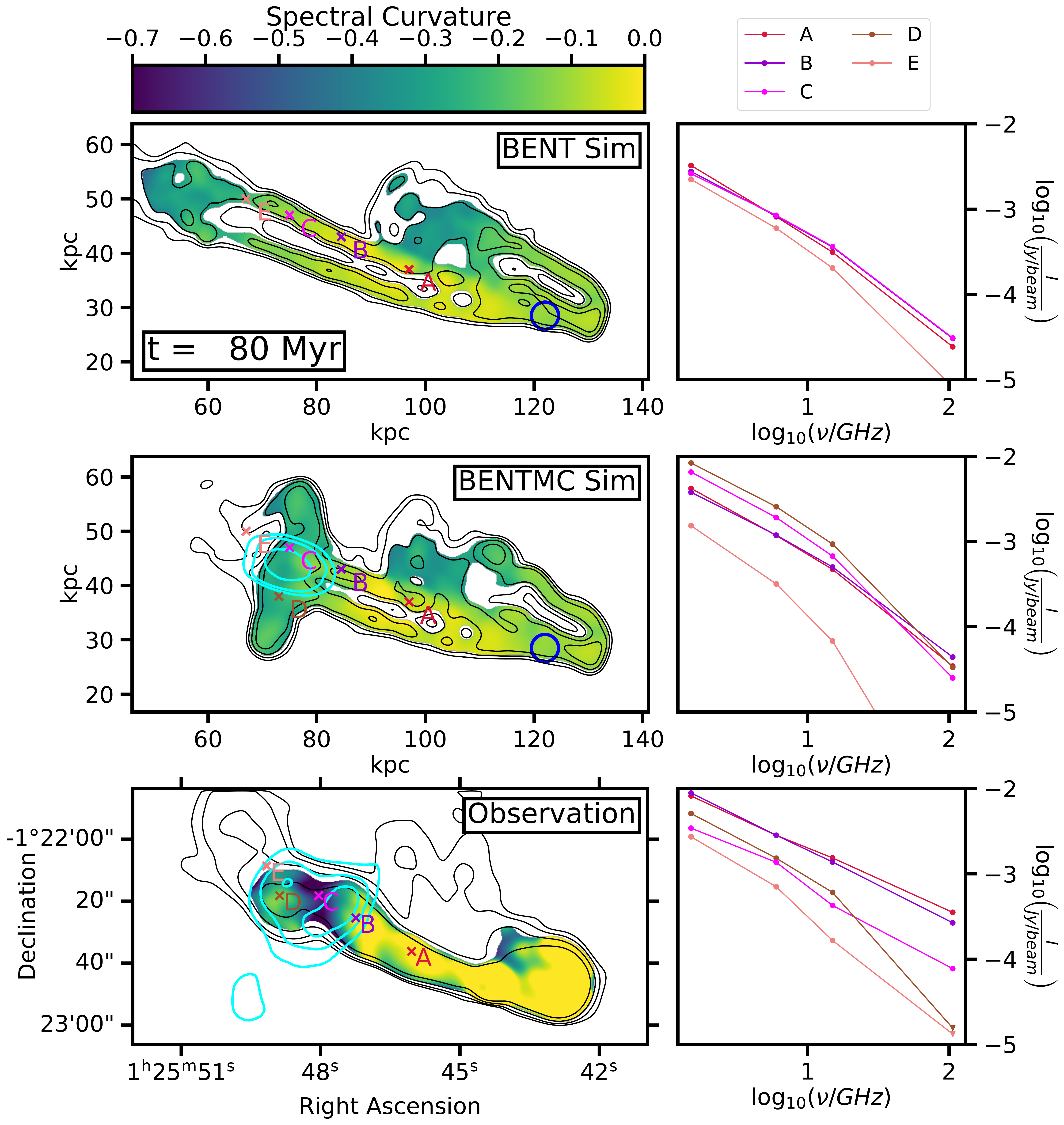}
    \caption{Radio spectral curvature (see Eq. \ref{eq:curvature}; 1.5--6--15\,GHz, cut at $150\mu Jy/\text{beam}$ at 15GHz) maps with 6\,GHz radio brightness contours (levels = [150, 300, 600, 1200] $\mu Jy/\text{beam}$ in black) for the \textbf{BENT} (top panels) and \textbf{BENTMC} (middle panels) simulations and for NGC\,541 (bottom panels). Cyan contours are as in Figure \ref{fig:specIndexCombined}. Point spectra are plotted between 1.5--106\,GHz in the right panels for the locations labeled in the curvature maps. The 106GHz points in the bottom panel at points D and E are upper limits for detection. In the simulated images, the blue circle marks the location of the jet source. The beam sizes in each panel and cutout size in the bottom panel are the same as in Figure \ref{fig:specIndexCombined}. An animated version of this figure (10 sec) displaying the evolution for the left panels of this figure from $t=0$ to 80 Myr is available online. We exclude the point spectra from the movie as it is difficult to define dynamically equivalent points of interest throughout the simulation time. Again, the \textbf{BENTMC} simulation begins to deviate from what is seen in \textbf{BENT} once the cloud starts interacting with the jet (at $t\approx 45$ Myr),  as jet material begins to rebound laterally away from the interaction site and radiatively ages as it is slowly advected with the wind. \url{https://www.youtube.com/watch?v=r6sylSpYtAI}}
    \label{fig:curvature}
\end{figure}

The left panels of Figure \ref{fig:curvature} show spectral curvature maps for the \textbf{BENT} and \textbf{BENTMC} simulations between 1.4, 6.0 and 15\,GHz as well as for NGC\,541. For NGC\,541, we combined the C- and Ku-band data described in Section \ref{obs} with L-band (1.4\,GHz) continuum data from \citet{lacy:alma}; we also use the ALMA Band 3 (106\,GHz) continuum data from that paper for the point spectra in the right-hand panels of Figure \ref{fig:curvature}. In NGC\,541, the eastern jet shows little curvature until it reaches MO, where it curves down significantly. In the deflection region, the spectrum is less negatively curved than near the center of MO. The curvature in the simulations undergoes less abrupt changes. In the \textbf{BENT} simulation, the spectrum shows little curvature until the jets reach the point where turbulence dominates and they transition to become less collimated ``tails,'' which show a more negatively curved spectrum. This is also true in the \textbf{BENTMC} simulation up until the interaction point. Then, stronger negative curvature occurs closer to the jet source due to the presence of the cloud disrupting the forward propagation of the jet. Similar to the case of NGC\,541, the deflected portion of the jet from \textbf{BENTMC} has a slightly less curved spectrum than material nearer to the dense cloud core. 
In the simulations, this curvature comes from radiative aging of the plasma as it piles up in front of the dense cloud. However, in the deflection region near MO the spectrum is less curved than near the core of MO. This may be due to emission mechanisms not captured in the simulations (see the discussion at the end of Section \ref{index}).

In the panels on the right side of Figure \ref{fig:curvature}, spectra are plotted for specific labeled points from the associated left panels. In NGC\,541, these points roughly correspond to: (A) In the core of the jet well before the interaction with MO, (B) the core of the jet just upstream of MO, (C) emission near the center of MO, (D) in the deflected region of the jet, and (E) emission from the jet just past MO. Simulated points in \textbf{BENTMC} were chosen to match these regions dynamically from the simulated images, but follow along the bright edge of the jet rather than the low radio surface brightness core. Points from \textbf{BENT} are the exact same points as from \textbf{BENTMC} but without point (D) since there is no deflection present. Examining the point spectra between \textbf{BENT} and \textbf{BENTMC}, it is clear that the presence of the cloud has a significant effect. In \textbf{BENT}, without a cloud, the spectrum remains relatively constant from points A-C and only begins to show signs of aging and spectral curvature as the jet reaches point (E) where jet flapping has disrupted the collimation. However, \textbf{BENTMC} shows an enhancement of surface brightness at points (C) and (D) at low frequencies and also introduces significant curvature, which is quite extreme at point (E). Some of these same trends are also seen in NGC\,541: there is little to no change between point (A) and (B); point (C) is highly curved between 1.5 - 15GHz, though seems to flatten again at high frequency; and the intensity drops off at point (E).  

\subsection{Interaction Timescale} \label{timescales}

From the evidence presented in Section \ref{bentJets}, we have established that the \textbf{BENTMC} simulation is the best match for the radio observables from MO. Assuming this simulation reflects the dynamics of the NGC 541--MO system, we can estimate an age for the encounter based on the simulation time from the cloud's first contact with the jet until the time that the simulation most resembles the radio morphology, spectral, and polarization signatures of MO that we have described. In the simulation, the cloud interaction begins around 45\,Myr into the simulation. Significant deflection of the jet material is visible in the projected viewing angle by about $t = 55$\,Myr. The change in polarization angle at the location of the cloud occurs around this time as well, as it is a result of the jet's deflection. Around $t = 60$\,Myr, the core of the cloud enters fully into the jet and we observe deflection laterally in all directions away from the interaction point rather than a flow deflected in one direction. This is different than the observed deflection of the NGC\,541 jet at MO, which appears to deflect in only one direction. This may mean that MO is either not yet fully aligned with the core of the jet, or more likely, that its trajectory has a slight offset (impact parameter) such that it is a more glancing collision. The timing of this will depend on the velocity of the cloud, but from the values chosen for this study, we estimate the interaction age of MO to be on the order of $\tau \sim 10-20$\,Myr. 

This interaction duration is qualitatively well-matched with the stellar population age analysis of C06 ($\sim 7$ Myr), suggesting that little star formation took place within the HI cloud associated with MO prior to the interaction with the radio jet. With new HST WFC3/UVIS UV-Optical and WFC3/IR NIR observations of MO we will clarify this picture by identifying all massive star clusters as well as performing pixel-by-pixel ($\sim 0.1"$) spectral energy distribution fitting using simple stellar populations (SSP; Linden et al. in prep). Future simulations will also reexamine the dynamical scenario of \textbf{BENTMC}, but with the physics of star formation and the requisite cooling and chemistry included, akin to F17. The simulations presented in that paper included a front of current star formation and a secondary burst further downstream, surrounded by older stars that formed a few Myr earlier in the interaction. New simulations would be required to see how this may differ in this new dynamical picture.

\section{Conclusions}\label{conclusions}

We have presented new VLA observations, analyzed along with archival data, to create a more complete picture at radio wavelengths of the jet from NGC\,541 and its interaction with Minkowski's Object. To augment our investigation, we have also performed a series of numerical simulations that self consistently model the MHD plasma and cosmic ray electron populations to enable calculation of synthetic radio maps. By performing the same radio analyses with the VLA and simulated data, we are able to directly compare observation to theory and find strong support for a specific interaction scenario. Our main conclusions are:

\begin{enumerate}
    \item NGC\,541 is a bent tail jet source that is highly projected relative to our line of sight. Simulations that do not include bent jets are unable to match the morphology of NGC\,541 and MO.
    
    \item The eastern jet of NGC\,541 is clearly interacting with MO. The strongest pieces of evidence for this are:
    
    \begin{enumerate}
    \item The apparent (projected) trajectory of the eastern jet changes direction just past the center of MO. This may represent deflection, which is a major feature of our jet-cloud-interaction simulations.
    \item The radio spectral index steepens at the location of MO. In our simulations, this occurs due to piling up of jet plasma in the interaction region, which then ages radiatively.
    \item The polarization angle abruptly changes at the location of MO, with vectors rotated $\sim 90^\circ$ with respect to the deflected jet trajectory. The interaction of a cloud with the jet in our simulation produces similar changes in polarization vector orientation and may imply that the jet in NGC\,541 remains well collimated as it deflects away from the interaction point. 
    \item The radio spectrum becomes strongly negatively curved at the location of MO as jet material radiatively ages. In the deflection region the jet emission is less curved, suggesting the importance of emission from stars and supernovae in MO due to jet-induced star formation.
   
  \end{enumerate}
  
\end{enumerate}

Improving our understanding of Minkowski's Object and other cases of jet-induced star formation will require expanding and extending the observation-informed jet simulation framework presented here. For MO, constraints from additional ALMA CO observations with sufficient surface brightness sensitivity to identify and characterize diffuse, low-surface brightness gas are needed. At the same time, increasing the number of candidate analogs to MO, particularly at high redshift, is essential. Joint observations with the VLA and ALMA have the potential to do this. However, observing limitations (e.g. sensitivity and survey speed) pose challenges to 
building large statistical samples. Finally, new and future radio/millimeter instruments and surveys, such as the next-generation Very Large Array \citep{nyland+18}, CMB-S4 \citep{abazajian+19}, DSA-2000 \citep{hallinan+19}, and the Square Kilometre Array \citep{braun+15}, 
will provide systematic searches for positive jet-driven feedback through cosmic time. 

\acknowledgments

Support for program \#HST-GO-14722.001-A was provided by NASA through a grant from the Space Telescope Science Institute, which is operated by the Associations of Universities for Research in Astronomy, Incorporated, under NASA contract NAS5-26555. This work used the Extreme Science and Engineering Discovery Environment (XSEDE), which is supported by National Science Foundation grant number ACI-1548562. We gratefully acknowledge the support of the National Science Foundation through grants AST-1907850 (CN and PCF) and PHY-1748958 (PCF). The National Radio Astronomy Observatory is a facility of the National Science Foundation operated under cooperative agreement by Associated Universities, Inc. Basic research in radio astronomy at the U.S. Naval Research Laboratory is supported by 6.1 Base Funding. We thank Sarah Wood for assistance with the processing some of the VLA data used in this paper.

\vspace{5mm}
\facilities{The simulations presented here were run and analyzed at the College of Charleston on their high performance Linux cluster (https://hpc.cofc.edu)}

\software{Astropy \citep{2013A&A...558A..33A},  
          Cloudy \citep{2013RMxAA..49..137F}, 
          SExtractor \citep{1996A&AS..117..393B},
          Wombat \citep{wombat}
          }

\bibliography{minkypol}{}
\bibliographystyle{aasjournal}

\end{document}